\documentclass[a4paper]{article} %,twocolumn %11pt,

\usepackage{arxiv} 

\usepackage{siunitx}
\usepackage{subcaption}
\usepackage[british]{babel}%[french,british]{babel} % Dernière langue = Langue principale %,english US
\usepackage{hyperref}

\usepackage{graphicx}
\usepackage[dvipsnames]{xcolor}

\usepackage[utf8]{inputenc} % Required for inputting international characters 
\usepackage[T1]{fontenc} % Output font encoding for international characters 
\usepackage{mathpazo} % Use the Palatino font by default 
\usepackage{amsfonts}
\usepackage{amsmath}

\usepackage[backend=bibtex,%bibtex,%biber
			maxnames=3,minnames=3,%
			style = ieee,%phys,%
			citestyle=numeric-comp,%
			sorting=none,%
			date=iso,%iso% ou ymd
			urldate=iso,%
			seconds=true
			]{biblatex} %Biblatex
\title{Thermal characterisation by Scanning Photothermal Radiometry using a random undersampled measurement scheme}% Your title
\author{
 Florian Crouau\textsuperscript{1,2},  Alejandro Mateos-Canseco\textsuperscript{1,2}, Jérémie Maire\textsuperscript{1,2}, Jean-Luc Battaglia\textsuperscript{1,2}, Stéphane Chevalier\textsuperscript{1,2} \\ %ID ORCID 
  \textsuperscript{1} Univ. Bordeaux, CNRS, Bordeaux INP, I2M, UMR 5295\\
  \textsuperscript{2}  Arts et Metiers Institute of Technology, CNRS, Bordeaux INP, I2M, UMR 5295\\
  351 Cours de la Libération, F-33400 Talence \\
  \texttt{jeremie.maire@u-bordeaux.fr} }

\date{ 2026--05--15 }% Your date

\begin{document}

% Create the title of the paper
\maketitle 

% write your abstract in the `abstract` environment
\begin{abstract}
Scanning Photothermal Radiometry (SPR) is an active thermal technique that is simultaneously non-destructive, contactless, and allows for temporal resolutions on the order of nanoseconds, spatial resolutions down to the sub-micrometre scale and at different depths. This scanning method can be time consuming thus this work shows that it is possible to reduce the amount of measurements taken by 6 when using SPR on a sample consisting of carbon fibres in an aluminium matrix. It uses irregular sampling on sparse signals, and a weighted random technique to further decrease the amount of samples needed. 
\end{abstract}

\section{Introduction}

Photothermal non destructive testing consists in a wide array of methods to distinguish materials based on their thermal properties deduced from their infrared emission or re-emission. These methods exists at the macro scale such as flash thermography, but also at the micro scale with photothermal radiometry \cite{salazarSizingDepthWidth2022, hamaouiSpatiallyLocalizedMeasurement2020, zengMeasurementHightemperatureThermophysical2021}. A micro scale imaging technique based on photothermal radiometry, called Scanning Photothermal Radiometry (SPR) has been developed recently and allows precise measurements with spatial resolutions reported at \SI{0.5}{\um} \cite{mateos-cansecoThermalCharacterizationVertical2024} in some cases along the plane. SPR can also gather information about the material at a range of depths by modulating the input signal frequency. This leads to 3 dimensional data which contains a lot of information about the sample. However the full acquisition can become time consuming as each point has to be measured at every frequency, and often contain redundant information.

SPR\cite{mateos-cansecoThermalImagingScanning2023} is an imaging technique derived from Photothermal Radiometry, and more specifically Modulated Photothermal Radiometry (MPtR)\cite{nordalPhotothermalRadiometry1979}. It allows for the measurement of 3D photothermal images with both in plane and depth information. This technique can be used for precise spatial measurements that are limited by the minimum step size of the scanning motor, size of the laser spot and of the detector, which have been proved to detect defects at the micro and sub micrometer scale \cite{mateos-cansecoThermalCharacterizationVertical2024, hamaouiSpatiallyLocalizedMeasurement2020}. At this scale, issues arise related to the compromise between acquisition times and noise levels for a given image field of view. Thus, under-sampling techniques have been studied in similar scan imaging, especially with the rise of CS \cite{candesNearOptimalSignalRecovery2006, donohoCompressedSensing2006}, such as in Frequency-Domain Thermoreflectance \cite{yangThermalpropertyMicroscopyCompressivesensing2025}, Scanning Trasmission Electron Microscopes \cite{kovarikImplementingAccurateRapid2016}, or Atomic Force Microscopy \cite{niuFastAFMImaging2021}.

Thus this work explores under sampling techniques applied to SPR in order to diminish this drawback. To do so, a random sampling scheme is used at a given frequency followed by algorithmic reconstruction of the full image. Two reconstruction techniques are compared here: Compressive Sensing (CS) \cite{candesStableSignalRecovery2005} using a cosine transform, and Radial Basis Function (RBF) interpolation \cite{hardyMultiquadricEquationsTopography1971}. In order to find which technique is more suited for SPR, the comparison is done on two main criterion that are the noise level and the sub sampling ratio. Finally a new sampling scheme is proposed in order to scan materials at multiple depths efficiently.  

This article starts with a description the SPR experimental setup, a brief explanation of the two methods of reconstructing images from randomly undersampled set of points that are used here, then we propose a sampling scheme based on a first measurement. The two reconstruction methods are compared in terms of robustness to low sampling ratio and high noise. Finally the proposed method is used with the most efficient reconstruction method in order to acquire a 3D set of images by sweeping frequencies.

\section{Methods}

\subsection{Scanning Photothermal Radiometry}

\begin{figure}
\centering 
\includegraphics[width = \textwidth]{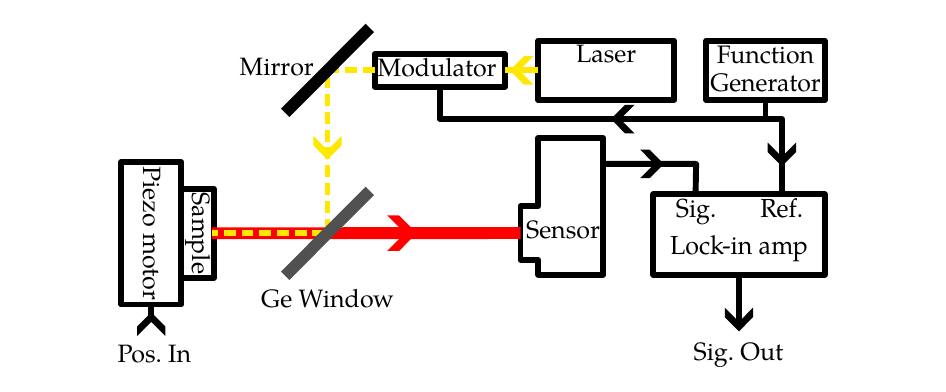}

\caption[SPR: Setup]{Schematic of Scanning Photothermal Radiometry Setup. A laser at \SI{1064}{\nm} focuses a beam of waist \SI{3}{\um} on the sample after going through an acousto-optic modulator (AOM) and being reflected by a Germanium window. The material response is sent back through the window to the detector. A lock-in then extracts the signal at the frequency given on the input by a waveform generator to the AOM. The measure is then repeated for every chosen frequency and position by controlling a piezo-electric motor on which the sample is placed. } \label{fig:SPR_Setup}
\end{figure} 

%TODO Déplaçable vers l'intro, après Paragraphe 2 ? 
Scanning Photothermal Radiometry (SPR), as Photothermal Radiometry relies on the use of infrared (IR) detectors to measure the emission of a material under a localized photothermal excitation. The emittance ($M$) is the IR radiative response to this excitation, and is described by the Stefan-Boltzmann law, such that $M = \epsilon \sigma T^4$, where $T$ is the temperature of the material, $\epsilon$ the emissivity coefficient of the surface and $\sigma$ the Stefan-Boltzmann constant. As the variation $\Delta T$ induced by the excitation is usually very small and for a periodic input signal at frequency $f$ as is the case in Modulated Photothermal Radiometry (MPtR), the variation of emittance can be linearised as $\Delta M = 4 \epsilon \sigma T_0^3 \Delta T_f \cos(2\pi f t + \varphi) $. Thus, measuring through a lock-in amplifier the amplitude of $\Delta M$ relative to the modulated input signal at frequency $f$ gives a value proportional to $\Delta T_f$, while the measured phase offset gives $\varphi$. Both amplitude and phase are dependant on the thermal properties of the studied material and thus both can be used for their estimation \cite{coronaRecentProgressModulated2023}. %for testing ultimately and imaging. 
However they each have their own drawbacks: On one hand, the amplitude depends on the emissivity $\epsilon$ of the sample surface which makes the study of inhomogeneous materials more limited such as the composite tested here. On the other hand while the phase is independent from $\epsilon$ and all else being equal, its noise level tends to be higher.

The experiment setup is as described in \autoref{fig:SPR_Setup} and in previous works \cite{mateos-cansecoThermalImagingScanning2023}. A laser at \SI{1064}{\nm} is modulated by an acousto-optic modulator (AOM) at a given frequency $f$ in the range \qtyrange{1}{100}{\kHz}. It is focused with a beam waist of $r = \SI{3}{\um} $ on the sample thanks to a parabolic mirror, and reflected toward the sample thanks to a Germanium window. The output signal at IR wavelengths transmitted by the Germanium window is collected by a detector with a wider field of view of \SI{11}{\um}. A Lock-in amplifier then extracts the amplitude $A$ at frequency $f$ and phase shift $\varphi$ relative to the modulation. The measurement is then repeated in order to create an image by scanning the sample along both axes thanks to a piezoelectric translation stage. In the regular sampling case, the step size is \SI{1.5}{\um} and defines the pixel size in the image. Finally each point is measured at a given set of chosen frequencies in order to retrieve information about various depths of the material as the frequency is inversely proportional to the square root of the diffusion length, thus leading to a set of maps at multiple depths. 

The two methods are compared on a given ROI of a carbon fibre in an aluminium matrix. The sample was scanned on a $40\times 40$ grid of pixel size \SI{1.5}{\um}, with a lock-in frequency at \SI{6}{\kHz}. The acquisition time per measured point is limited by the lock-in time constant $\tau = \SI{30}{\ms}$, as a settling time of $10\tau$ is required, with additional delays caused by communication speed with the motor and data saving. Which leads to an acquisition time on the order of $1-2$ seconds per point per frequency, thus a full $40\times 40$ image takes \SI{30}{\min} to \SI{1}{h}. A study on a wider field of view and at multiple depths can become time consuming, thus a faster acquisition is desirable.

\subsection{Compressive Sensing}

Compressive Sensing (CS) is a technique that allows the reconstruction of signals with less measurements that what would be required by the sampling theorem thanks to an added hypothesis of sparsity and non-regular sampling. A signal is said to be sparse when it can be described by only a few non-zeros values in a given basis. Here the discrete cosine transform (DCT) space is used. For $y$ the measurement vector of size $M$ and $x$ the ground truth of size $N$ such that $N>M$ and with $\mathbf{H}$ a $N\times M$ matrix representing the undersampling, the direct problem can be written simply as: 

\begin{equation}
y = \mathbf{H}x
\end{equation}

The goal of CS is to solve the inverse problem, by assuming that it is possible to write $x$ in another basis such that $x = \mathbf{\Psi} \xi$ where $\xi$ is mostly comprised of zeros. First, the basis $\mathbf{\Psi}$ must be correctly chosen such that the sparse hypothesis is true. Here the 2D DCT basis is chosen due to the fact that it is an universal basis \cite{bruntonDataDrivenScienceEngineering2019} that can be calculated efficiently. Then, the amount of non-zero values need to be minimised, which can be achieved by the use of a norm called ``$\ell_0$-norm'' that effectively counts the non-zeros in a vector. This pseudo-norm is computationally difficult to calculate, thus a panel of approaches have been proposed in the literature from thresholding to deep learning \cite{yuanSnapshotCompressiveImaging2021}, with the closest convex norm often used to approximate it. The inverse problem is therefore a $\ell_1$-norm minimisation such that: 

\begin{equation}
\hat{\xi} = \min_\xi ||\mathbf{H\Psi} \xi - y||_2 + \lambda ||\xi||_1
\end{equation} This inversion requires a regularisation parameter $\lambda$ that will depend on how sparse the signal is. The reconstruction is then its inverse transform $\hat{x} = \mathbf{\Psi}\hat{\xi}$. This minimisation is made thanks to the OWL-QN (Orthant-Wise Limited-memory Quasi-Newton) algorithm \cite{andrewScalableTrainingL1regularized2007,taylorCompressedSensingPython}, that was shown in previous work \cite{crouauUndersampledFlyingSpot2026} to be able to efficiently compute this minimisation. 

\subsection{Radial Basis Functions Interpolation}

Radial Basis Functions (RBF) interpolation is one of the main meshfree method working under the conditions of unstructured and multidimensional data, and is related to Kriging \cite{fazioSpatialInterpolationAnalytical2013} also called Gaussian process regression which is the best linear unbiased estimation for this type of problem. Its objective is to find $x$ the ground truth from a set of measurements $y$ at known positions $u$ (i.e. $y_i = x(u_i)$) by solving a problem of the form: 

\begin{equation}
y = \begin{pmatrix} 
A_{11} & A_{12} & \cdots & A_{1N} \\
A_{21} & A_{22} & \cdots & A_{2N} \\
\vdots  & \vdots  & \ddots & \vdots  \\
A_{N1} & A_{N2} & \cdots & A_{NN} \\
\end{pmatrix}  \begin{pmatrix}
w_1  \\ w_2  \\ \vdots \\ w_N
\end{pmatrix}\label{eq:RBFInterp}
\end{equation}
With $y$ the measurement vector, $w$ the vector of weights to be determined, and the matrix $\mathbf{A}$ containing elements $A_{ij}$ that are radial basis function chosen appropriately. These RBF are a class of function that depend only on the distance between the points $i$ and $j$ of positions $u_i$ and $u_j$, such that $A_{ij} = A(||u_{i} - u_{j}||_2)$. Here, the thin plate spline $r^2 \log(r)$, with $r$ the radial distance between $u_{i}$ and $u_{j}$, is used because it does not require any tweaking of the scale factor, while still leading to satisfactory results. 

The weights $w$ can be calculated by inverting $\mathbf{A}$, $w =\mathbf{A}^{-1}y$, as $\mathbf{A}$ can be shown to be non-singular as long as the measurement points are distinct \cite{micchelliInterpolationScatteredData1986}. It is then possible to calculate $x(u) = \sum_i^N w_i A(||u - u_{i}||)$ at any position $u$. 

For high noise levels, a regularised, ie. smoothed, RBF reconstruction is made by adding a parameter $\lambda$ such that $y = (\mathbf{A} + \lambda \mathbf{I})w$ \cite[p.167]{fasshauerMeshfreeApproximationMethods2007}. This reconstruction relaxes the conditions by allowing the result to not strictly pass through the data points, it is therefore not an interpolation but instead an approximation. The result image is similar to the non regularised one smoothed via a convolution product. This RBF approximation has additional drawbacks by adding a bias and requiring the correct choice of regularisation parameter $\lambda$ similarly to the OWL-QN case. The RBF interpolation and approximation algorithm are implemented in SciPy \cite{fasshauerMeshfreeApproximationMethods2007,RBFInterpolatorSciPy}.

\subsection{Random and weighted random samplings}

\begin{figure}
\centering 
\includegraphics[width = \textwidth]{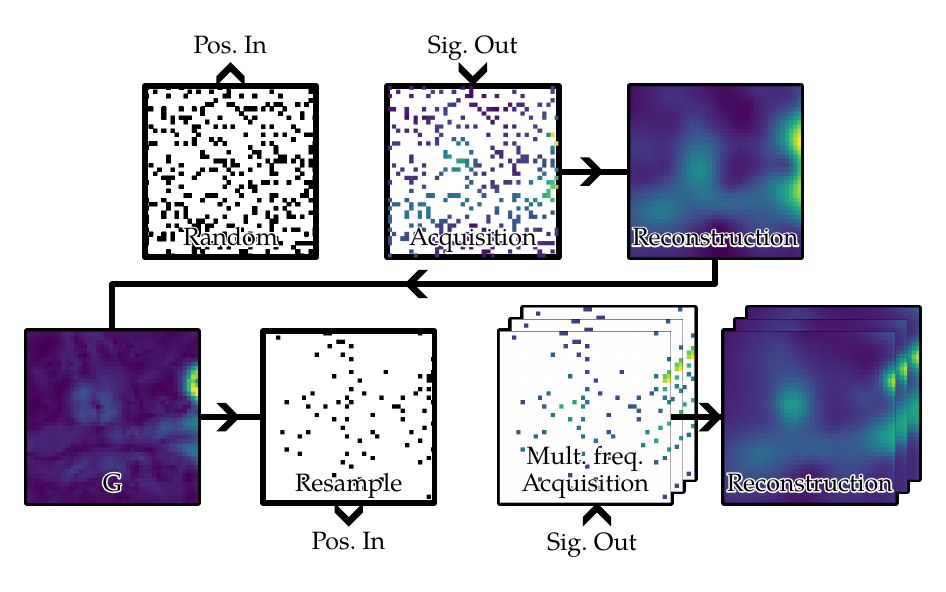}

\caption[SPR: Weighted Random Principle]{Sub-sampling scheme at multiple frequencies, consisting in 2 parts: A sub-sampled acquisition (top) at a given frequency, followed by a stronger sub-sampling at all the other frequencies (bottom). First a fully random series of points is created, each point acting as a set of coordinates for the piezo controller, then the associated signal at those points is measured thanks to the SPR setup \autoref{fig:SPR_Setup}, from these acquisitions at known positions, reconstructions algorithms are used to approximate the equivalent full image. The gradient of the image is calculated and used in order to sample preferably where the gradient is high. This new sampling is then used again on the setup but this time along all the frequencies.} \label{fig:SPR_PrincipWeigh}
\end{figure} 

The regular scan in both space and frequencies can quickly become time consuming, usually taking hours, thus a weighted random scheme was developed and implemented to further optimise the acquisition time beyond the purely random sampling, as described in \autoref{fig:SPR_PrincipWeigh}. First, a fully random sampling is used at a first frequency, in the case presented here $25\%$ of a $40\times 40$ image at $f=\SI{6}{\kHz}$, then the full image $I$ is reconstructed. Its gradient is calculated along both axes and used as a probability density. This probability proposed here is defined per pixel as: %
\begin{equation}
P_i = \frac{ G_i + \alpha\max (G) }{ \sum_i ( G_i + \alpha\max (G) )}
\end{equation} %
with $G_i = |\frac{\partial I}{\partial x}| + |\frac{\partial I}{\partial y}|$  and $\alpha $ an arbitrary value set at $\alpha = 0.02$ here in order to still have a few samples in the homogeneous regions of the image. Then samples are drawn according to these probabilities with a number of samples inferior than in the fully random case. Measurements are then made according to this re-sampling where a range of frequencies are acquired for each point. The norm $\ell_1$ is used to define $G_i$, as the objective is to create probabilities proportional to the gradient, an $\ell_2$ norm would also work with marginal differences.  
The underlying assumption behind this re-sampling is that overall the material remains similar when probed at higher frequencies. In the case of the sample studied here of carbon fibre in an aluminium matrix, the fibres are along the depth. A likely spot for defects is at the interface between the fibre and the matrix, so the new sampling is made to focus on this part as the gradient at interfaces is higher than within a fibre for instance and is even greater when a defect is present.

\section{Results}

\begin{figure}
\centering 

\begin{subfigure}{0.24\linewidth} 
	\centering 
	\includegraphics[width = \textwidth]{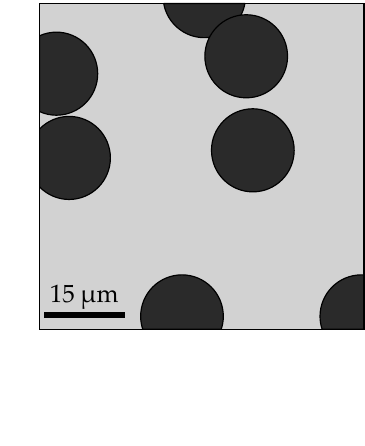}
	\caption{Schematic} \label{fig:SPR_Schematic}
\end{subfigure}
\hfill
\begin{subfigure}{0.24\linewidth} 
	\centering 
	\includegraphics[width = \textwidth]{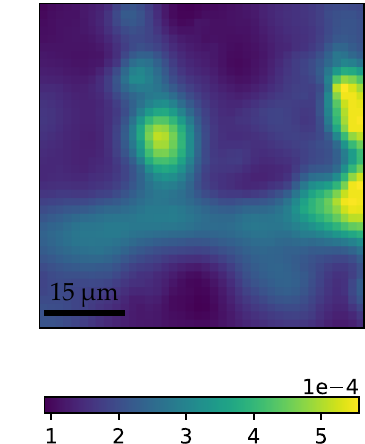}
	\caption{Amplitude (u.a.)} \label{fig:SPR_Amplitude}
\end{subfigure}
\hfill
\begin{subfigure}{0.24\linewidth} %[c]
	\centering 
	\includegraphics[width = \textwidth]{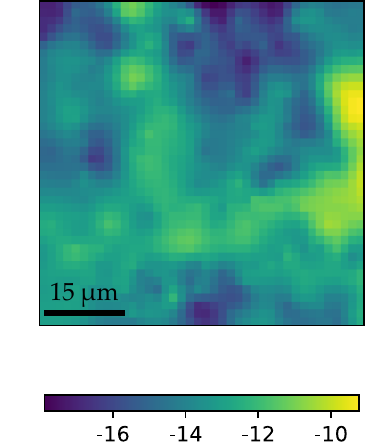}
	\caption{Phase ($\deg$)} \label{fig:SPR_Phase}
\end{subfigure}

\caption[SPR: Sample]{Schematic of the sample \subref{fig:SPR_Schematic}, and its response at \SI{6}{\kHz} in amplitude \subref{fig:SPR_Amplitude} and phase \subref{fig:SPR_Phase} (Reconstruction from $40\%$ of the samples, input laser power of \SI{1}{\W} and using RBF). The pixel pitch corresponds to a displacement of \SI{1.5}{\um}. The sample is comprised of an aluminium matrix (Grey) and carbon fibres of \qtyrange{12}{15}{\um} (Black).} \label{fig:SPR_Sample}
\end{figure}

Two criterion are studied for the amplitude: Robustness to the number of samples, i.e. sampling ratio, and robustness to noise. A schematic of the ROI of the sample is drawn in \autoref{fig:SPR_Schematic}, it contains carbon fibres in an aluminium matrix. The fibres are around \SI{15}{\um} in diameter and can be distinguished from the background as plateaux of amplitudes in \autoref{fig:SPR_Amplitude}. High variations in amplitude can correspond to defects such as detachment between fibre and matrix or porosities within the matrix but also can be caused by the difference of thermal conductivity between the matrix and the fibre. 

\subsection{Comparison on samplings ratios at a given frequency}

\begin{figure}
\centering 
\includegraphics[width = \textwidth]{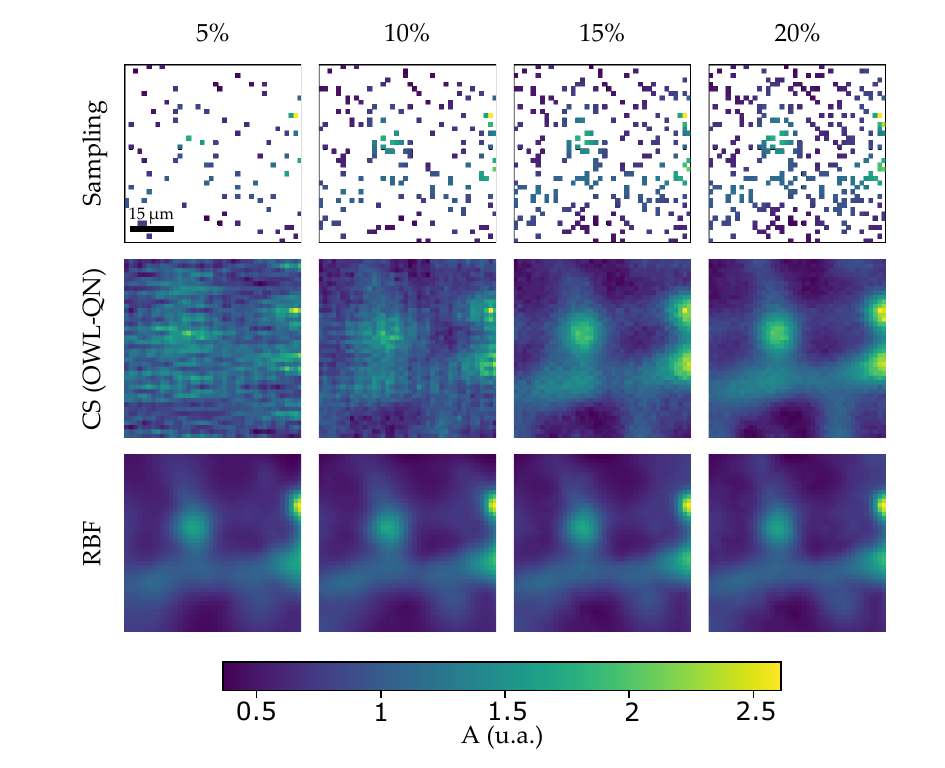}

\caption[SPR: Comparison (Sampling Ratios)]{Reconstruction results for a $40 \times 40$ image with the two methods for different sampling ratios at a laser power of \SI{750}{\mW}. The first line contains the samples used for reconstruction in the other two lines. These samples are taken randomly at a given ratio, from $5\%$ to $20\%$, compared to the full image. They are also such that each increase of samples always includes the points from the previous image. For the OWL-QN algorithm, the regularisation parameter is set to $0.1$. } \label{fig:SPR_SampRatios}
\end{figure} 

The dependence on the number of samples is studied first, with a laser power of \SI{1}{\W}. The number of samples is progressively decreased from $25\%$ (400 samples) to $5\%$ (80 samples) with each case being a subset of the previous one. \autoref{fig:SPR_SampRatios} shows for each column the undersampled acquisition, followed by the reconstructions both with the CS and RBF methods.% A relative error graph is provided in \autoref{fig:SPR_Samp_graphRMSE} calculated as a root mean squared between the result and the reconstruction at $25\%$ using RBF.

 As expected reconstruction quality decreases as the number of samples decreases. For low sampling ratios, the RBF interpolation clearly outperforms the OWL-QN reconstruction. Even a number of samples as low as $5\%$ suffice to reconstruct the image with minimal loss of quality in RBF, while this limit is at roughly $15\%$ for OWL-QN. Overall, this difference is general but the actual limit ratio will depend on the measured structure. Here, the image is quite smooth due to the fact that the step size is smaller than the beam waist of the laser, therefore there is a convolutive effect between successive points which is a favourable scenario for undersampling. 
 
 \subsection{Comparison on noise levels at a given frequency}

\begin{figure}
\centering 
\includegraphics[width = \textwidth]{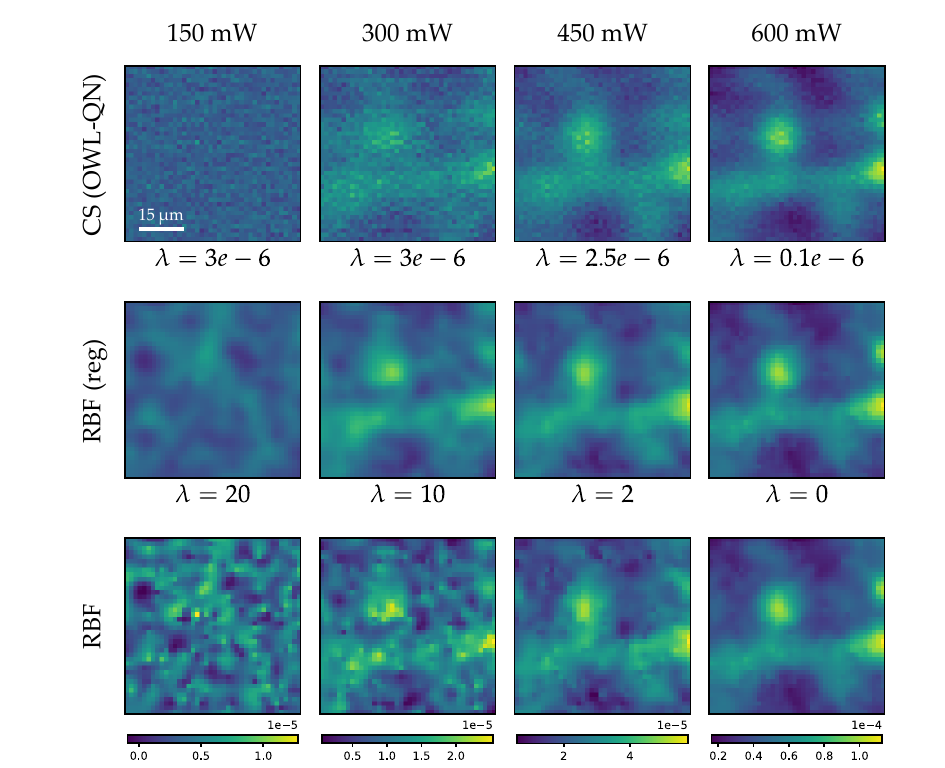}

\caption[SPR: Comparison (Noise levels)]{Reconstruction results for a $40 \times 40$ image taken at \SI{6}{\kHz} with the 3 methods (in lines) for different laser powers (in columns) ranging from \SI{150}{\mW} to \SI{600}{\mW} at a random sampling ratio of $25\%$.} \label{fig:SPR_Noise}
\end{figure}

The robustness to noise was also studied. Successive images were made with decreasing laser input power, which increases the relative noise levels accordingly. The sampling is exactly the same in every image, with a ratio of $25\%$. The results for the range \SI{600}{\mW} to \SI{150}{\mW} are shown in \autoref{fig:SPR_Noise}. The OWL-QN result has a better visual fidelity compared to the interpolation result especially at \SI{300}{\mW} and  \SI{450}{\mW}, where it is easier to distinguish the structure although the range of values is lowered. However, by using by adding a regularisation parameter to the RBF, the resulting approximation manages to give acceptable resulting images at \SI{300}{\mW} at the expense of an added bias to the values of amplitude. At the highest noise levels, ie. lowest power of \SI{150}{\mW}, both methods are unable to reconstruct the signal, thus the images are not exploitable.  

Reconstructions were illustrated here with the amplitude signal. The phase also contains a lot of information in SPR and thus could be treated similarly. However, the phase images are noisier which, as shown in \autoref{fig:SPR_Noise}, has a detrimental impact on the reconstruction results and thus would require a modification of the experimental parameters, such as the laser power, lock-in time constant or by averaging multiple acquisitions. 
In other terms, the relative time gain between a phase image and its undersampled equivalent is the same as this is determined by the sampling ratio, but as they both require slower acquisition speed the minimum achievable time will be greater than for images of amplitude. 

\subsection{Multiple frequencies scan}

\begin{figure}
\centering 

%\rotatebox[origin=l]{90}{\makebox[2ex]{OWL-QN}}
\begin{subfigure}{0.24\linewidth} 
	\centering 
	\includegraphics[width = \textwidth]{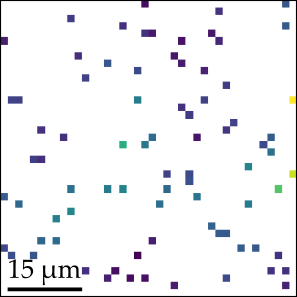}
	\caption{Random sampling} \label{fig:SPR_Rand_Samp}
\end{subfigure}
\hfill
\begin{subfigure}{0.24\linewidth} 
	\centering 
	\includegraphics[width = \textwidth]{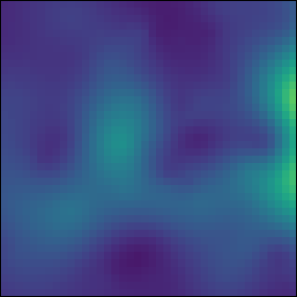}
	\caption{Random \SI{1}{\kHz}} \label{fig:SPR_Rand_1}
\end{subfigure}
\hfill
\begin{subfigure}{0.24\linewidth} %[c]
	\centering 
	\includegraphics[width = \textwidth]{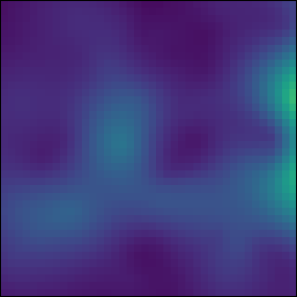}
	\caption{Random \SI{40}{\kHz}} \label{fig:SPR_Rand_2}
\end{subfigure}
\hfill
\begin{subfigure}{0.24\linewidth} %[c]
	\centering 
	\includegraphics[width = \textwidth]{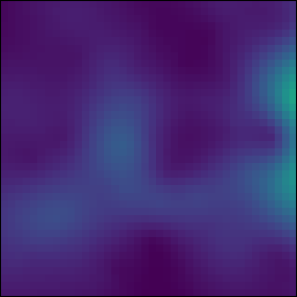}
	\caption{Random \SI{100}{\kHz}} \label{fig:SPR_Rand_3}
\end{subfigure}

%\rotatebox[origin=l]{90}{\makebox[2ex]{RBF}}
\begin{subfigure}{0.24\linewidth} 
	\centering 
	\includegraphics[width = \textwidth]{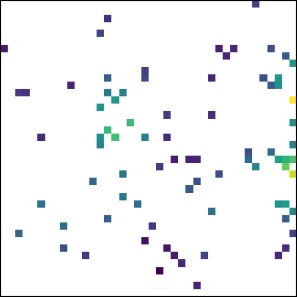}
	\caption{Weighted sampling} \label{fig:SPR_WeighRand_Samp}
\end{subfigure}
\hfill
\begin{subfigure}{0.24\linewidth} 
	\centering 
	\includegraphics[width = \textwidth]{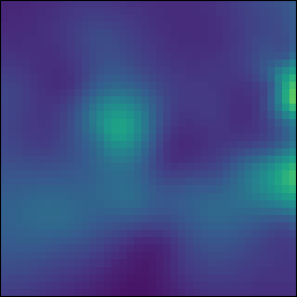}
	\caption{Weighted \SI{1}{\kHz}} \label{fig:SPR_WeighRand_1}
\end{subfigure}
\hfill
\begin{subfigure}{0.24\linewidth} %[c]
	\centering 
	\includegraphics[width = \textwidth]{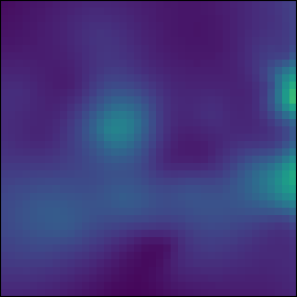}
	\caption{Weighted \SI{40}{\kHz}} \label{fig:SPR_WeighRand_2}
\end{subfigure}
\hfill
\begin{subfigure}{0.24\linewidth} %[c]
	\centering 
	\includegraphics[width = \textwidth]{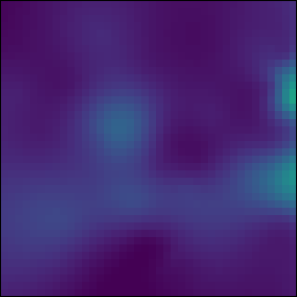}
	\caption{Weighted \SI{100}{\kHz}} \label{fig:SPR_WeighRand_3}
\end{subfigure}

\caption[SPR: Weighted random]{Reconstruction results for a $40 \times 40$ image at multiple frequencies.} \label{fig:SPR_WeighRand}
\end{figure}

As described in the methods section, a weighted random resampling is deduced from the previous reconstruction at \SI{6}{\kHz}, and then used to scan the sample at a set of frequencies in the range \qtyrange{1}{100}{\kHz}. These images are compared to the fully random sampling used previously at the same sampling ratio of $5\%$ and a laser power of \SI{750}{\mW} in \autoref{fig:SPR_WeighRand}. Both lead to acceptable results in the given case, although the weighted random method tends to a give slightly better contrast.

Ultimately, both the minimum sampling ratio and the choice of the coefficient $\alpha$ in the probability to draw samples depends on the material and the type of defects that the user expects. A low value of $\alpha$ means that the entire sample is supposed to be similar and therefore can lead to the non detection of defects such as the apparition of cracks at a specific given depth, especially if said depth is further than the one associated with the first frequency of the calculated gradient. Thus it is primordial in NDT to carefully chose correct hypotheses as under sampling necessarily creates a compromise between acquisition time and fidelity. The weighted random sampling can easily be modified by changing the probability function in order to better represent other scenarios. For instance in the case studied here, if the objective was to look for defects within the aluminium, the probability could be defined as the inverse of what is shown here, ie. be maximum when the gradient is minimum. 

\section{Conclusion and Perspectives}

SPR can largely be accelerated thanks to algorithmic reconstruction of images from a small subset of samples. Both RBF interpolation and CS have been used to accurately represent images with $1/6$ of the samples that would have been required for a traditional sampling scheme, and thus an equivalent reduction in time, going from \SI{1}{h} to \SI{10}{\min} for one frequency. RBF interpolation led to overall better results with lower sampling ratios (as low as $1/20$) while still performing well when confronted to noisy data. The new proposed sampling scheme drawing samples according to the image gradient in order to scan a wide array of frequencies and therefore depths opens the way to a further reduction in acquisition time.

Further improvements can be achieved by applying the undersampling not only on the spatial scanning but also on the frequency signal, as successive frequencies are strongly correlated. Considerations on the photothermal PSF were outside of the scope of this article, but a thorough study of it is beneficial for a more precise localisation of defects as was proposed by Seidel \textit{et al.}\cite{seidelAttemptQuantitativePhotothermal1995} in one of the first attempt at photothermal imaging.

\subsection{References}

% Add your bibliography style here
%\bibliography{refs}

%\bibliography{Superres}

\end{document}